\documentclass[letterpaper, 10 pt, conference]{ieeeconf}%

\IEEEoverridecommandlockouts%
\usepackage{amsmath}%
\usepackage{amssymb}%
\usepackage{cite}%
\usepackage{algorithm}
\usepackage{algpseudocode}
\usepackage{multicol}
\usepackage{graphicx}
\usepackage{caption}
\usepackage{subcaption}
\usepackage{svg}
\usepackage{amsmath}
\usepackage{algorithm}
\usepackage{algpseudocode}

\usepackage{amsthm}

\newcommand{\normQ}[2]{\left\lVert #1 \right\rVert_{#2}^2}

\newtheorem*{remark}{Remark}

\title{\LARGE \bf
Paying for Space: Incentive-Aware  Motion Planning
for \\ Multi-Agent Collision Avoidance}

\author{Debajyoti Chakrabarti and Anushri Dixit%
\thanks{Debajyoti Chakrabarti is a PhD student in Aerospace Engineering,
Mechanical and Aerospace Engineering Department,
University of California, Los Angeles, CA 90095, USA
        {\tt\small debjyoti@ucla.edu}}%
\thanks{Anushri Dixit is an Assistant Professor in the Mechanical and Aerospace Engineering Department,
University of California, Los Angeles, CA 90095, USA
        {\tt\small anushridixit@ucla.edu}}%
}

\begin{document}

\maketitle
\thispagestyle{empty}
\pagestyle{empty}

\begin{abstract}

Advanced Air Mobility (AAM) systems require scalable coordination mechanisms to manage large fleets of aerial vehicles operating in shared, capacity-limited airspace. In such environments, different operators may have private preferences over trajectory characteristics, such as travel time, fuel consumption, or deviation from nominal routes. If centralized traffic management relies on self-reported preferences, operators may strategically misreport their costs to obtain more favorable trajectories.
This paper proposes a multistage motion planning framework augmented with mechanism design to enable collision avoidance for AAM systems with privately known costs. The proposed approach integrates convex safe corridor construction with a VCG-inspired mechanism to ensure conflict-free passage through constrained airspace while incentivizing truthful revelation of private preferences. Simulation results demonstrate safe and decentralized coordination among agents with heterogeneous preferences.

\end{abstract}

\section{INTRODUCTION}
Advanced Air Mobility (AAM) will require coordination of many autonomous aerial vehicles operating simultaneously in shared, capacity-limited airspace, where safety, efficiency, and access to constrained regions must be managed jointly \cite{ma2025deep,wu2025managing,kochenderfer2012next}. In such settings, trajectory planning is not only a geometric and dynamical feasibility problem, but also an allocation problem over scarce airspace resources. Different operators may have private preferences over travel time, control effort, fuel usage, or deviation from nominal routes, and may strategically misreport these preferences if a central planner relies on self-reported costs or constraints \cite{chen2025two,vickrey1961counterspeculation,clarke1971multipart,groves1973incentives}. To address this strategic aspect, one needs not only a collision-free allocation rule, but also a transfer mechanism that accounts for the externality one agent imposes on the others. Vickrey--Clarke--Groves (VCG) mechanisms provide the classical foundation for such externality-based allocation in quasi-linear settings \cite{nisan2007algorithmic}.

\begin{figure}[t]
    \centering
    \includegraphics[width=\columnwidth]{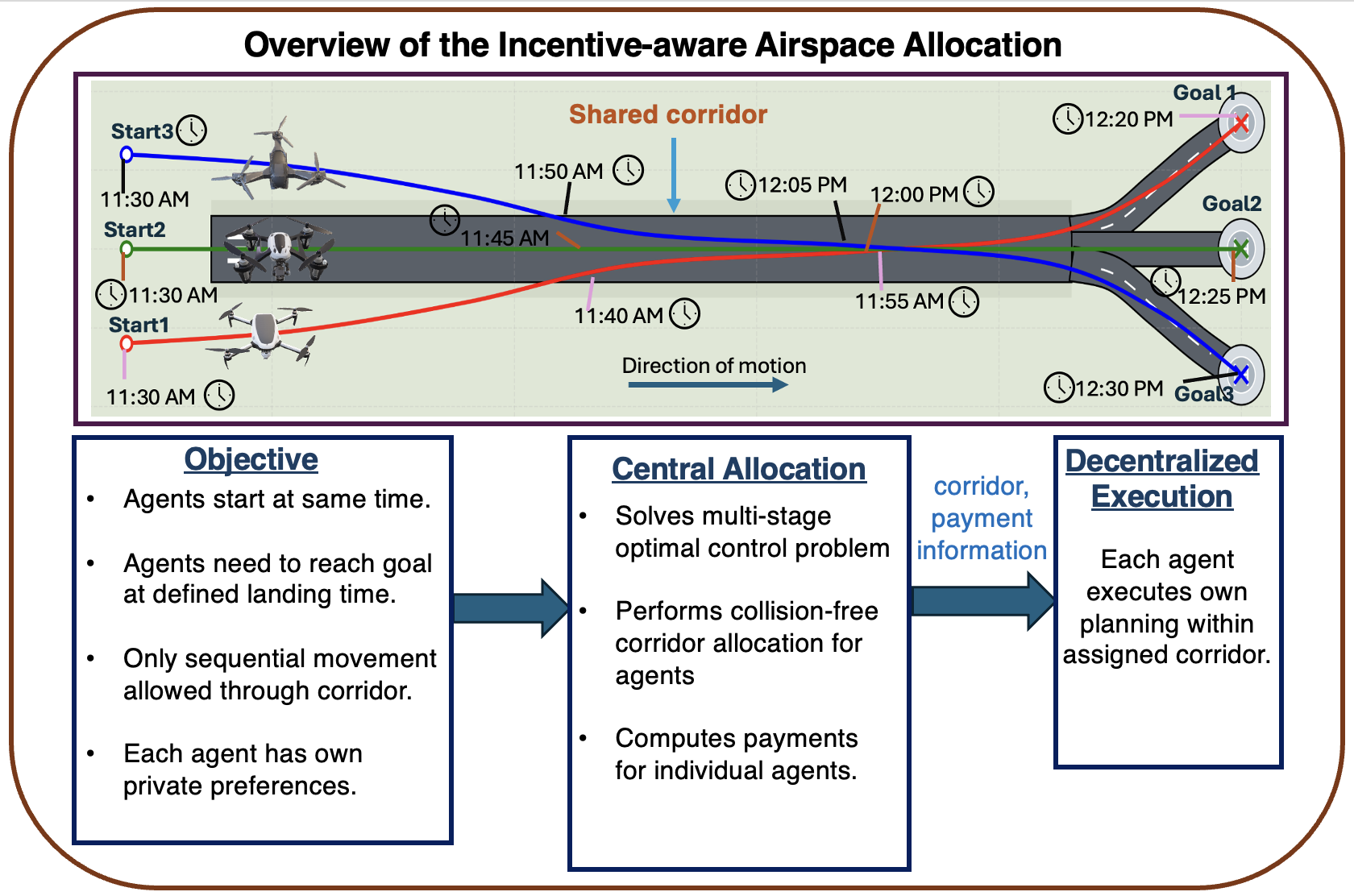}
\caption{An overview of the incentive-aware  collision free airspace allocation framework for AAM systems.}
    \label{fig:anchor}
\end{figure}

A large body of prior work studies multi-agent trajectory planning through search-based planning, sampling-based planning, reciprocal collision avoidance, optimal control, and safe trajectory generation \cite{orthey2023sampling,karaman2011sampling,richards2002aircraft,van2008reciprocal,mellinger2011minimum}. In parallel, safe multi-agent coordination has been addressed using reachability, control barrier functions, model predictive control, and game-theoretic safety formulations \cite{chen2016multi,wang2016safety,alonso2015collision}. These approaches have substantially advanced collision avoidance and dynamic feasibility, but they typically assume cooperative agents or truthfully specified objectives and therefore do not directly address the incentive issues that arise when agents have privately known costs. Even without strategic behavior, safety becomes challenging when many vehicles interact in shared constrained airspace: recent work has emphasized that pairwise safety guarantees do not, in general, characterize the true multi-agent safe set because of conflicting constraints, and ``leaky corner'' effects \cite{choi2025resolving}. Related recent work has also highlighted the role of uncertainty-aware prediction and conformal planning for safe interaction among dynamic agents \cite{dixit2023adaptive,stamouli2024recursively}. At the system level, recent studies in AAM and air traffic management continue to emphasize the need for scalable coordination for flow management, congestion mitigation, structured airspace allocation \cite{ma2025deep,wu2025managing, wu2025optimization}.

Motivated by these gaps, this paper develops a \emph{VCG-based}, incentive-aware framework for collision-free space allocation with privately known agent preferences. The proposed approach combines a VCG-inspired transfer structure with convex corridor allocation based on pairwise separating hyperplanes and continuous mixing variables, enabling non-overlapping safe corridors without mixed-integer scheduling of agents. A bilevel formulation is used to connect centralized allocation with decentralized execution, and a multistage decomposition is introduced to obtain a tractable approximation. In contrast to prior work on strategyproof trajectory planning with local decision-making power \cite{chen2025two} and prior work on multi-agent safety without private-information incentives \cite{choi2025resolving,dixit2023adaptive}, our framework studies collision-aware air corridor allocation with a VCG-based payment layer and decentralized realizability. Rather than claiming exact dominant-strategy incentive compatibility for the final approximate implementation, we use the VCG structure to design an incentive-aware allocation rule and evaluate its behavior empirically through truthful-versus-misreport simulations. Fig.~\ref{fig:anchor} illustrates the proposed incentive-aware pipeline
for collision-free space allocation among agents.

The main contributions of this paper are as follows. First, we propose an incentive-aware, VCG-based framework for multi-agent air corridor allocation under collision-avoidance constraints. Second, we formulate the problem as a bilevel allocation-and-execution model that explicitly connects centralized airspace allocation with decentralized agent-level trajectory generation, ensuring consistency between the centrally computed allocation and the local optimization solved by individual agents. We provide a tractable multistage decomposition of the bilevel optimization. Third, we introduce a convex safe-corridor construction based on pairwise
separating hyperplanes with optimized locations, enabling non-overlapping
corridor allocation. Finally, we validate the proposed framework through simulation studies that assess nominal performance and empirical incentive behavior under truthful and misreported preference reports.

The remainder of the paper is organized as follows. Section II reviews the VCG mechanism in cost form and introduces the notation used throughout the paper. Section III formulates the collision-aware air corridor allocation problem and presents the bilevel formulation and its tractable multistage decomposition. Sections IV and V present simulation setup and results, followed by conclusions and future work.

\section{Preliminaries: VCG in Cost Form}

Consider $N$ agents indexed by $i\in\{1,\dots,N\}$. Let
$y\in\mathcal{Y}$ denote an allocation from a report-independent feasible
set, with private type $\theta_i$ and reported type $\tilde{\theta}_i$ for
agent $i$. In cost form, the VCG allocation
\cite{vickrey1961counterspeculation,clarke1971multipart,
groves1973incentives,nisan2007algorithmic} is,
\begin{equation}
y^\star(\tilde{\theta})
\in
\arg\min_{y\in\mathcal{Y}}
\sum_{i=1}^{N}J_i(y,\tilde{\theta}_i).
\label{eq:vcgalloc}
\end{equation}

The realized quasilinear utility of agent $i$ is
\begin{equation}
u_i
=
-J_i(y^\star(\tilde{\theta}),\theta_i)
-t_i(\tilde{\theta}),
\label{utileq}
\end{equation}
where the Clarke-pivot transfer is
\begin{equation}
\begin{aligned}
t_i(\tilde{\theta})
&=
\sum_{j\neq i}
J_j(y^\star(\tilde{\theta}),\tilde{\theta}_j)\\
&\quad-
\min_{y\in\mathcal{Y}_{-i}}
\sum_{j\neq i}
J_j(y,\tilde{\theta}_j).
\end{aligned}
\label{ti1}
\end{equation}
Here $\mathcal{Y}_{-i}$ denotes the feasible allocation problem with agent
$i$ removed. Under the convention in~\eqref{utileq}, $t_i>0$ denotes a
charge paid by agent $i$, while $t_i<0$ denotes a credit or subsidy.

For the exact allocation~\eqref{eq:vcgalloc}, if the feasible set
$\mathcal{Y}$ is independent of the reported types, truthful reporting
$\tilde{\theta}_i=\theta_i$ is a weakly dominant strategy under the standard
VCG assumptions \cite{nisan2007algorithmic,makowski1987vickrey}.
\section{Problem Formulation}

We now formulate the multi-agent motion planning problem for AAM using the VCG framework. We consider a scenario where a central planner (like an automated air traffic controller) allocates individual, collision-free zones to multiple aerial agents operating in a shared airspace that are expected to land at specified times, subject to collision avoidance and dynamical constraints. Each agent has private preferences encoded through cost matrices. A direct application of VCG  requires the planner to compute an airspace allocation that minimizes the sum of costs (with reported preferences) over a feasible set of collision-free paths.

However, in our application, the execution of an assigned plan is decentralized: each agent ultimately plans its own trajectory by solving an individual optimal control problem. Therefore, it is not sufficient for the centralized solution to be optimal for the combined cost; it must also be individually optimal for the agents under decentralized execution. In other words, each agent has its own objective, and it is not sufficient for a centralized planner to combine all the objectives of different agents into a single weighted sum objective using the weights or costs that are (mis)reported.

To formulate this multi-agent motion planning problem, we construct a bilevel optimization framework with: 1) a lower-level optimization that computes each
agent's best motion plan within its collision-free airspace, 2) an upper-level optimization that determines the collision-free airspace
allocation minimizing the overall cost of all agents. We now formalize this
bilevel framework.

\subsection{Central Allocation Problem}
\noindent Consider $N$ aerial agents with the discrete-time dynamics:
\begin{equation*}
    x_{i,k+1} = A_ix_{i,k} + B_iu_{i,k},
\end{equation*}
where, each agent $i\in\{1,\dotsc, N\}$ has state $x_{i,k} \in \mathcal{X} \subseteq \mathbb{R}^{n_x}$, and control input $u_{i,k} \in \mathcal{U} \subseteq \mathbb{R}^{n_u}$ at time $k \in \{0 ,\dotsc, T \}$ with reported types $\tilde{\theta}_i$. The linear dynamics are computed using, $A_i\in\mathbb{R}^{n_x \times n_x}$ and $B_i\in \mathbb{R}^{n_x \times n_u}$. We combine all the states of agent $i$ for all time $k \in \{0 ,\dotsc, T \}$,
\[
X_i = \{x_{i,k}\}_{k=0}^{T}, \quad
U_i = \{u_{i,k}\}_{k=0}^{T-1},
\]
and the all agent states are combined as,
\[
X = \{X_i\}_{i=1}^N, \quad
U = \{U_i\}_{i=1}^N.
\]

Our goal is to assign each agent $i$ some operating free space at every time $k$, $\mathcal{P}^{\text{free}}_{i,k}$, that has no overlap with the free space assigned to other agents $j\neq i$. In this work, we limit the parameterization of the free space to convex polytopes, $\tilde{\mathcal{P}}^{\text{free}}_{i,k}(\mu_{i,k})$, where we parameterize the edges of $\mathcal{P}^{\text{free}}_{i,k}$ using the variable $\mu_{i,k}$,
\[
\mu_{i,k} = \{\mu_{ij,k}\}_{j \neq i}, \quad
M_i = \{\mu_{i,k}\}_{k=0}^{T}, \quad
M = \{M_i\}_{i=1}^N.
\]
This $\mu_{i,k}$ encodes the location of the hyperplane that separates agents $i$ and $j$ at any time $k$. Henceforth, we call $\mu$ a \textit{mixing variable}.

{\remark{As an illustrative example, consider the line passing through the position of agent $i$ and perpendicular to the line segment connecting agents $i$ and $j$ is given by $Ax=b_i$. Similarly, the line passing through the position of agent $j$ is given by $Ax=b_j$. If $x_j \not\in \{x: Ax\leq b_i\}$, and $x_i \not\in \{x: Ax\geq b_j\}$, any line $Ax =  b_i + \mu_i(b_j-b_i)$ will separate $x_i$ and $x_j, \forall \mu_i\in[0,1]$. We can find such separating hyperplanes, parameterized by $\mu$, between all agents in a pairwise manner and for all time $k$. }}

Moreover, the free space polytope, $\tilde{\mathcal{P}}^{\text{free}}_{i,k}(\mu_{i,k})$, assigned to every agent $i$ at time $k$ with size parameterized by $\mu_{i,k}$, must satisfy,
\begin{align}
\tilde{\mathcal{P}}^{\text{free}}_{i,k}(\mu_{i,k}) &\subseteq \mathcal{P}^{\text{free}}_{i,k}, \label{eq7}\\
\tilde{\mathcal{P}}^{\text{free}}_{i,k}(\mu_{i,k}) \cap 
\tilde{\mathcal{P}}^{\text{free}}_{j,k}(\mu_{j,k}) &= \emptyset, \quad i \neq j, \label{eq8}\\
\bigcup_{i=1}^{N} \tilde{\mathcal{P}}^{\text{free}}_{i,k}(\mu_{i,k}) &= \mathcal{P}\subseteq\mathcal{X}, \label{eq9}
\end{align}
ensuring a conservative, non-overlapping partition of the feasible geometric space $\mathcal{P}$ across agents.

Using this parameterization of the collision-free region, the central planner solves the following bilevel optimal control problem, 
\begin{equation}
\begin{aligned}
\min_{X,U,M} \quad & \sum_{i=1}^{N} J_i(X_i, U_i, \tilde{\theta}_i) \\
\text{s.t.} \quad 
& p_{i,k} = \mathcal{C}x_{i,k}, \,\,p_{i,k} \in \mathcal{P}^{\text{free}}_{i,k}, \quad \forall i,k \\
& \tilde{\mathcal{P}}^{\text{free}}_{i,k}(\mu_{i,k}) \subseteq \mathcal{P}^{\text{free}}_{i,k}, \quad \forall i,k \\
&\tilde{\mathcal{P}}^{\text{free}}_{i,k}(\mu_{i,k}) \cap \tilde{\mathcal{P}}^{\text{free}}_{j,k}(\mu_{j,k}) = \emptyset, \quad i \neq j \\
&\bigcup_{i=1}^{N} \tilde{\mathcal{P}}^{\text{free}}_{i,k}(\mu_{i,k}) = \mathcal{P}, \\
& (X_i,U_i) = \arg\min_{\bar{X_i},\bar{U_i}} J_i(\bar{X}_i,\bar{U}_i,\tilde{\theta}_i), \qquad \forall i \\
&\qquad\qquad\qquad\text{s.t.} \quad \bar{x}_{i,k} \in \mathcal{X} \subseteq \mathbb{R}^{n_x}, 
\\ &\qquad\qquad\qquad\qquad u_{i,k} \in \mathcal{U} \subseteq \mathbb{R}^{n_u},\,\, \forall k, \\
& \qquad\qquad\qquad\qquad \bar{x}_{i,k+1} = A_i\bar{x}_{i,k} + B_i\bar{u}_{i,k}, \,\, \forall k, \\
&\qquad\qquad\qquad \qquad \bar{x}_{i,0} = x_i^{\text{init}}, \\
& \qquad\qquad\qquad\qquad \bar{x}_{i,k} = x_i^{\text{goal}},\; k \ge T_{\text{landing, i}}, \\
&\qquad\qquad\qquad\qquad \bar{p}_{i,k} = \mathcal{C}\bar{x}_{i,k}, \quad \forall k \\
&\qquad\qquad\qquad \qquad \bar{p}_{i,k} \in \tilde{\mathcal{P}}^{\text{free}}_{i,k}(\mu_{i,k}), \quad \forall k. \label{eq:central1}
\end{aligned}
\end{equation}
In~\eqref{eq:central1}, $T_{\text{landing},i}$ denotes the assigned landing time for agent $i$. Let $p_{i,k} = \mathcal{C}x_{i,k}$ and $\bar{p}_{i,k} = \mathcal{C}\bar{x}_{i,k}$ denote the position of agent $i$, where $\mathcal{C}$ is a projection operator extracting the position components of the state. The upper level of the bilevel program optimizes for the best split of the free space into time-varying polytopes assigned to each agent. The lower layer computes the optimal trajectory for each agent in a decentralized manner (as is the case at execution time).

We define the collision-free set for agent $i$ at time $k$ as
\begin{equation}
\mathcal{P}^{\text{free}}_{i,k}
=
\mathcal{P} \setminus \bigcup_{j \neq i} \mathbb{B}(p_{j,k}, r),
\end{equation}
where $\mathbb{B}(\cdot,r)$ denotes a closed ball of radius $r$ (each agent is assumed to have radius $r$). Equivalently, collision avoidance can be expressed as the pairwise separation constraint,
\begin{equation}
\|p_{i,k} - p_{j,k}\|_2 \geq 2r, \quad \forall j \neq i, \label{collision1}
\end{equation}
which enforces that agents remain separated at all times.

\subsection{Cost and Reports}

In this paper, we consider agent $i$ to have private parameters $\theta_i = (Q_i,Q_{f,i},R_i)$ (associated with stage, terminal and control costs, respectively) with $(Q_i, Q_{f,i}, R_i)\succeq 0$, and reports $\tilde{\theta}_i = (\hat Q_i, \hat Q_{f,i}, \hat R_i)$. Let
$\normQ{z}{Q}:=z^\top Q z$.
For a given goal state $x_i^{\text{goal}}$, the trajectory cost for any agent $i$, $i \in \{1,\dotsc,N\}$, is defined as,
\begin{eqnarray*}
J_i(X_i,U_i,\theta_i)
&=&
\sum_{k=0}^{T-1}
\left(
\normQ{x_{i,k}-x_i^{\text{goal}}}{Q_i}
+
\normQ{u_{i,k}}{R_i}
\right)
\nonumber \\
&&
+\normQ{x_{i,T}-x_i^{\text{goal}}}{Q_{f,i}}.
\end{eqnarray*}

\subsection{Convex Collision-free Set Construction}

The collision-free set $\mathcal{P}^{\text{free}}_{i,k}$ defined earlier is nonconvex due to pairwise exclusion constraints. To obtain a tractable formulation, we construct convex inner approximations of this set using a polytopic representation, building upon the approach described in~\cite{morgan2014scp}.

\subsubsection{Halfspace Library}

For each pair of agents $(i,j)$ with $i \neq j$ and time step $k$, we define the unit normal,
\begin{equation}
a_{i|j,k}
=
\frac{p_{j,k} - p_{i,k}}{\|p_{j,k} - p_{i,k}\|}, \qquad
a_{j|i,k} = -a_{i|j,k} \label{eq:normal}.
\end{equation}

\noindent We define the corresponding anchors as,
\begin{equation}
b_{i|j,k} = a_{i|j,k}^\top p_{j,k}, \quad
b_{j|i,k} = a_{j|i,k}^\top p_{i,k}, \label{eq:anchors}
\end{equation}
 and the associated tangent halfspaces are given by, 
\begin{eqnarray}
\mathcal{H}_{i|j,k}
=
\{p \in \mathbb{R}^{n_d} : a_{i|j,k}^\top p \geq b_{j|i,k} - r , \}\nonumber\\
\mathcal{H}_{j|i,k}
=
\{p \in \mathbb{R}^{n_d} : a_{i|j,k}^\top p \geq b_{i|j,k} - r , \}. \label{eq:halfspace}
\end{eqnarray}
For agent $i$, we stack the $(N-1)$ rows for all $j\neq i$ into
$A_{i,k}\in\mathbb{R}^{(N-1)\times n_d}$ and $b_{i,k}\in\mathbb{R}^{(N-1)}$. These halfspaces define separating hyperplanes between agents. The halfspaces separating an agent $i$ from other agents $\neg i$ form a convex polytope whose construction we describe next.

\subsubsection{Formulation of Separating Halfspaces}

We now use the mixing variables $\mu_{ij,k}\in[0,1]$ to parameterize the
allocation of separating space between agents. These variables satisfy
\begin{equation}
\mu_{ij,k}+\mu_{ji,k}=1,\qquad
\mu_{ii,k}=0,\quad \forall i\neq j.
\label{eq:mu-cons}
\end{equation}
For each pair $(i,j)$, the halfspace constraint at time $k$ is
\[
a_{i|j,k}^{\top}p_{i,k}
\leq
\mu_{ij,k}b_{i|j,k}
-
(1-\mu_{ij,k})b_{j|i,k}
-r.
\]
This represents a continuous allocation of the separating space between the
two agents.

Stacking all pairwise constraints yields the polyhedral set
\begin{equation}
\tilde{\mathcal{P}}^{\mathrm{free}}_{i,k}(\mu_{i,k})
=
\left\{
p\in\mathbb{R}^{n_d}:
A_{i,k}p
\leq
b_{i,k}(\mu_{i,k})-r\mathbf{1}
\right\}.
\label{eq:freespace_mu}
\end{equation}
By construction,
$\tilde{\mathcal{P}}^{\mathrm{free}}_{i,k}(\mu_{i,k})$
is a convex inner approximation of the collision-free region, with
$\mu_{ij,k}$ determining the allocation of separating space.

For a consistent pair satisfying~\eqref{eq:mu-cons}, the final halfspaces
place agents $i$ and $j$ on opposite sides of the same separating boundary
with margin $r$, yielding,
\begin{equation}
a_{i|j,k}^{\top}(p_{j,k}-p_{i,k})\geq 2r.
\label{eq:projected_separation}
\end{equation}
Since $\|a_{i|j,k}\|_2=1$, Cauchy--Schwarz gives,
\[
2r\leq
a_{i|j,k}^{\top}(p_{j,k}-p_{i,k})
\leq
\|p_{j,k}-p_{i,k}\|_2.
\]
Hence, whenever the final corridor-constrained problems are feasible,
$\|p_{j,k}-p_{i,k}\|_2\geq2r$, guaranteeing the required pairwise
collision separation.

\subsection{Multistage Decomposition of Central Allocation Problem}

The exact VCG allocation is strategyproof under the standard assumptions
when the reported social cost is minimized over a report-independent feasible
set. In our motion-planning setting,~\eqref{eq:central1} represents the
corresponding ideal allocation problem. Solving it directly is difficult
because the collision-avoidance constraints are nonconvex and bilevel
optimization is computationally challenging even for simpler problem
classes~\cite{salas2025bilevel}. We therefore use a three-stage decomposition
to obtain a tractable implementation. Since this decomposition need not recover
the global solution of~\eqref{eq:central1}, the standard VCG truthfulness
guarantee does not directly carry over. Accordingly, the implemented mechanism
is described as incentive-aware rather than strategyproof.

\textbf{In Stage 1}, we solve a centralized multi-agent trajectory optimization problem with explicit collision-avoidance constraints to generate nominal collision-free trajectories, that give us a feasible candidate for~\eqref{eq:central1}. These nominal trajectories are then used to construct pairwise separating halfspaces between agents at each time step.

\textbf{In Stage 2}, we solve decentralized agent-level individual problems to determine the halfspace mixing variables $\mu_{ij,k} \in [0,1]$, which determine how the shared corridor is split.
\begin{equation}
\begin{aligned}
(X_i,U_i, \mu_i) = \min_{X_i,U_i, \mu_i} \quad & J_i(X_i,U_i,\tilde\theta_i)\\
\text{s.t.} \quad
& {x}_{i,k+1} = A_i{x}_{i,k} + B_iu_{i,k},, \\
& x_{i,k} \in \mathcal{X}, \quad u_{i,k} \in \mathcal{U}, \,\,\forall k\\
& x_{i,0} = x_i^{\text{init}}, \\
& x_{i,k} = x_i^{\text{goal}}, \quad k \ge T_{\text{landing,i}}, \\
& {p}_{i,k} = \mathcal{C}{x}_{i,k}, \\
& {p}_{i,k} \in \tilde{\mathcal{P}}^{\text{free}}_{i,k}(\mu_{i,k}), \,\, \eqref{eq:mu-cons}, \label{agent2}
\end{aligned}
\end{equation}
The pairwise variables ($\mu_{ij},\mu_{ji}$) are then combined to enforce consistency between both agents, such that the updated values satisfy,
\[
\mu_{ij,k}^{\mathrm{new}} + \mu_{ji,k}^{\mathrm{new}} = 1.
\]
These combined mixing variables define the final polytopic corridor assigned to each agent. Because reconciliation modifies the independently optimized pairwise
boundaries, the preliminary trajectories need not remain feasible in the
reconciled corridors. The final corridor-constrained problems are therefore
resolved after reconciliation to verify feasibility.

\textbf{In Stage 3}, the free space corridors are fixed, and the resulting decentralized problem can now be separately solved for each agent, without any interdependence. Finally, we compute the VCG payments, using (\ref{ti1}), wherein~\eqref{eq:central1} must be solved without agent $i$. Hence, to compute $H_i^{\textrm{pivot}}$, we must resolve Stages 1-3 in the absence of agent $i$.

Once the planner solves the multistage decomposition and obtains the solution of the final stage, denoted by $y^*(\tilde{\theta}) = (X^*, U^*, M^*)$, it broadcasts to each agent its assigned feasible region $\tilde{\mathcal{P}}^{\text{free}}_{i,k}(\mu^*_{i,k})$ along with their scalar payments $t_i(\tilde{\theta})$.

The complete procedure is summarized in Algorithm~\ref{alg:hierarchical_vcg}.
\begin{algorithm}[th!]
\caption{Multistage VCG-Based Space Allocation}
\label{alg:hierarchical_vcg}
\footnotesize
\begin{algorithmic}[1]
\Require Agent reports $\tilde{\theta}=\{\tilde{\theta}_i\}_{i=1}^N$, central planner has the agents' initial states $\{x_i^{\mathrm{init}}\}_{i=1}^N$, goal states $\{x_i^{\mathrm{goal}}\}_{i=1}^N$, horizon $T_{\textrm{landing,i}}$
\Ensure Allocated safe corridors $\{\tilde{P}^{\mathrm{free}}_{i,k}\}$, and VCG transfers $\{t_i\}_{i=1}^N$

\State \textbf{Collect reports:} each agent submits $\tilde{\theta}_i$

\State \textbf{Stage 1: Centralized trajectory planning} \label{s1}
\State Solve a centralized multi-agent trajectory optimization with dynamics and collision-avoidance constraints and cost $\sum_{i=1}^{N}J_i(X_i, U_i, \tilde{\theta}_i)$ to obtain nominal trajectories $\{x_{i,k}^{\mathrm{nom}}\}$.

\For{$k=0,\dots,T_{\textrm{landing, i}}$}
    \For{each ordered pair $(i,j)$ with $i\neq j$}
        \State Construct halfspaces (\ref{eq:normal}), (\ref{eq:anchors}), and (\ref{eq:halfspace}).
    \EndFor
\EndFor

\State \textbf{Stage 2: Corridor allocation} \label{s2}
\For{each agent $i=1,\dots,N$}
    \State Solve~\eqref{agent2} for every agent $i$ to obtain $\mu_i$.
\EndFor

\State \textit{Mixing of pairwise allocations for consistency}:
\For{$k=0,\dots,T_{landing}$}
    \For{each unordered pair $\{i,j\}$ with $i\neq j$}
        \State Update mixing variables into one split,
        \[
        \mu_{ij,k}^{\mathrm{new}}
        = \frac{1}{2}\left(\mu_{ij,k} + 1 - \mu_{ji,k}\right),
        \quad
        \mu_{ji,k}^{\mathrm{new}} = 1 - \mu_{ij,k}^{\mathrm{new}}.
        \]
    \EndFor
\EndFor

\State \textit{Construct final shared corridors:}
\For{each agent $i=1,\dots,N$}
    \For{$k=0,\dots,T$}
        \State Form the allocated convex region
        \[
        \tilde{P}^{\mathrm{free}}_{i,k}(\mu_{i,k})
        =
        \left\{
        p \in \mathbb{R}^{n_d}
        \;:\;
        A_{i,k}p \le \tilde{b}_{i,k}(\mu^{\mathrm{new}}_{i,k} - r \mathbf{1}).
        \right\}
        \]
        where, \[
        \tilde{b}_{i,k}(\mu^{\mathrm{new}}_{i,k}= \mu_{ij,k} b_{i|j,k}
-
(1-\mu_{ij,k}) b_{j|i,k}.
        \]
    \EndFor
\EndFor \label{end2}
\footnotesize
\State \textbf{Stage 3: VCG transfer computation}
    \State Compute the Clarke-pivot transfer
    \[
t_i(\tilde{\theta})
=
\sum_{j\neq i}
J_j\!\left(y^*(\tilde{\theta}),\tilde{\theta}_j\right)
-
\sum_{j\neq i}
J_j\!\left(y^*_{-i}(\tilde{\theta}),\tilde{\theta}_j\right).
\]

\State \textbf{Return} $\{\tilde{P}^{\mathrm{free}}_{i,k}\}_{k=0}^{T_{landing, i}}$, $t_i(\tilde{\theta})$, $\forall i=\{1,\dotsc, N\}$.
\end{algorithmic}
\end{algorithm}
\subsection{Agent-level Decentralized Implementation}

With the information received from central planner, i.e., the allocated free space, $\tilde{\mathcal{P}}^{\text{free}}_{i,k}(\mu^*_{i,k}),\, \forall k$, each agent, $i$, then solves the decentralized optimal control problem,
\begin{equation}
\begin{aligned}
\min_{X_i,U_i} \quad & J_i(X_i,U_i,\theta_i)\\
\text{s.t.} \quad
& x_{i,k+1} = f(x_{i,k}) + g(x_{i,k})u_{i,k}, \\
& x_{i,k} \in \mathcal{X}, \quad u_{i,k} \in \mathcal{U}, \\
& x_{i,0} = x_i^{\text{init}}, \\
& x_{i,k} = x_i^{\text{goal}}, \quad k \ge T_{\text{landing,i}}, \\
& p_{i,k} \in \tilde{\mathcal{P}}^{\text{free}}_{i,k}(\mu^*_{i,k}), \quad \forall k \label{agent1}
\end{aligned}
\end{equation}

Since the transfer $t_i(\tilde{\theta})$ is independent of $(X_i,U_i)$, it does not affect the optimal solution. Therefore, the solution to~\eqref{agent1} is the same as the solution to, 
\begin{equation*}
\begin{aligned}
&\min_{X_i,U_i} \quad  J_i(X_i,U_i,\theta_i) + t_i(\tilde{\theta})\\
&\,\,\text{s.t. \,\quad constraints of (\ref{agent1})}.
\end{aligned}
\end{equation*}
which implies that the multistage airspace allocation problem also maximizes the agent-level return/utility, and conforms to the VCG mechanism-based incentive structure. 
Moreover, the centralized problem solution of the proposed framework is consistent with the solution of the decentralized agent-level problem within the assigned feasible regions.

\section{Results}

We consider an Advanced Air Mobility (AAM) landing-coordination problem in which multiple aerial agents must access a shared landing strip/corridor to reach their respective goals at the preassigned arrival times. 
\begin{figure}[t]
    \centering

    \begin{subfigure}[b]{0.75\columnwidth}
        \centering
        \includegraphics[width=\textwidth]{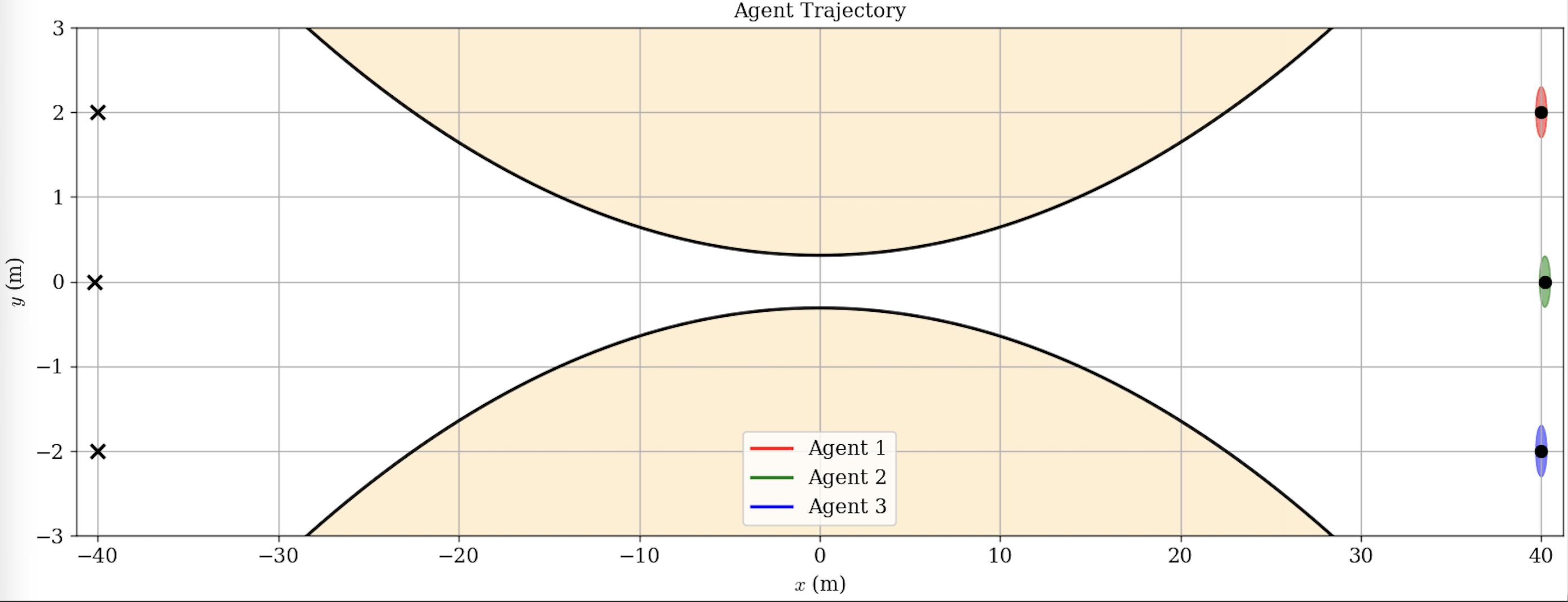}
        \caption{Trajectory begin}
        \label{fig:traj_begin}
    \end{subfigure}

    \vspace{0.5em}

    \begin{subfigure}[b]{0.75\columnwidth}
        \centering
        \includegraphics[width=\textwidth]{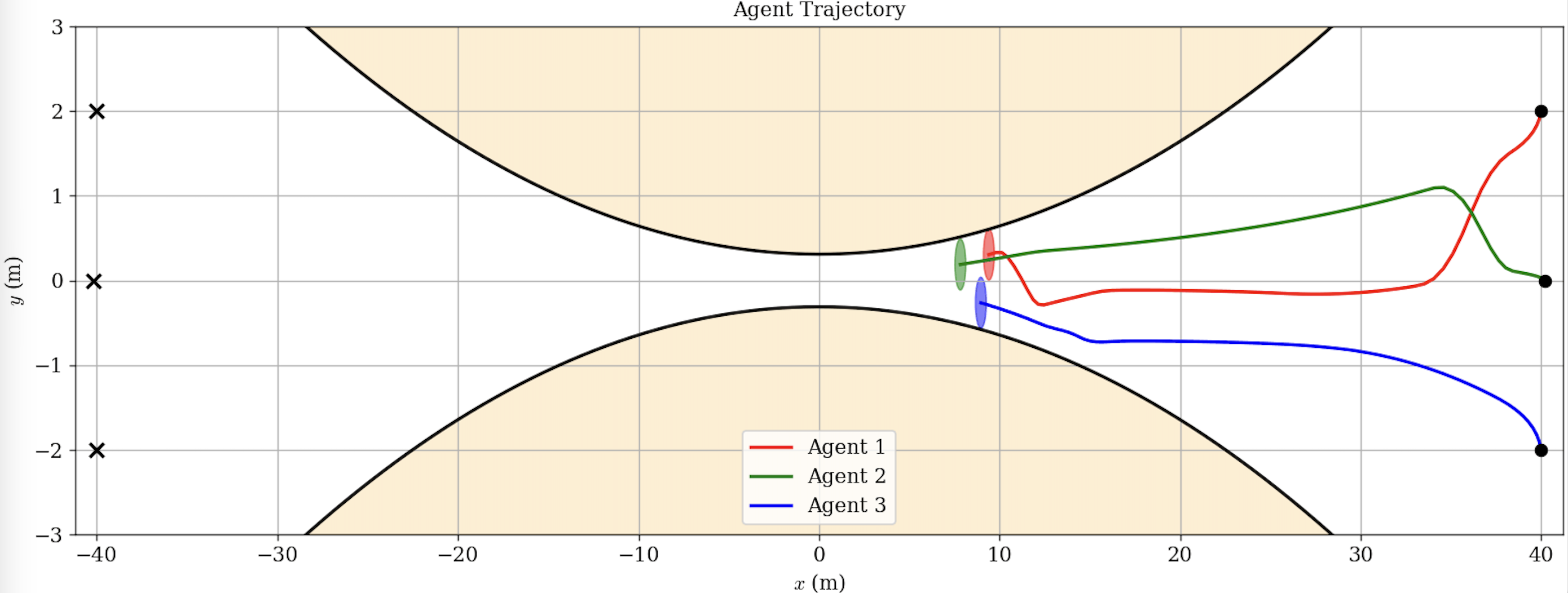}
        \caption{Tunnel entry}
        \label{fig:tunnel_entry}
    \end{subfigure}

    \vspace{0.5em}

    \begin{subfigure}[b]{0.75\columnwidth}
        \centering
        \includegraphics[width=\textwidth]{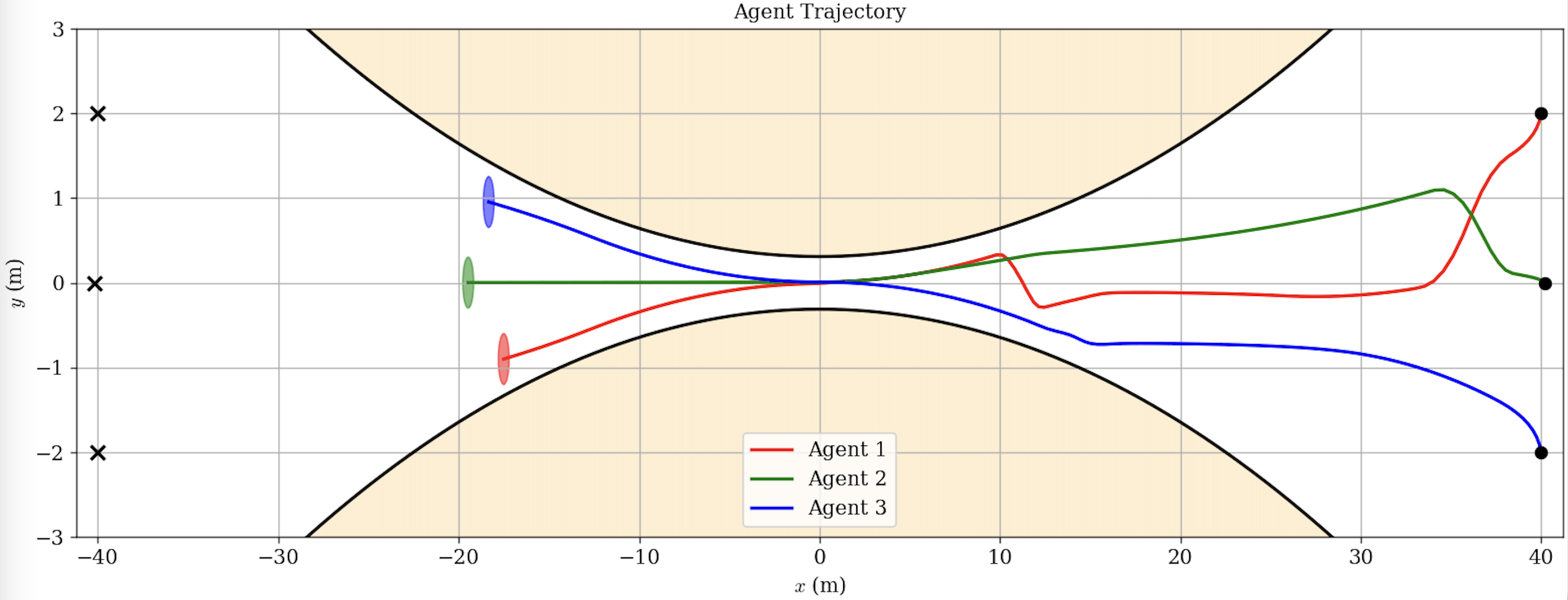}
        \caption{Tunnel exit}
        \label{fig:tunnel_exit}
    \end{subfigure}

    \vspace{0.5em}

    \begin{subfigure}[b]{0.75\columnwidth}
        \centering
        \includegraphics[width=\textwidth]{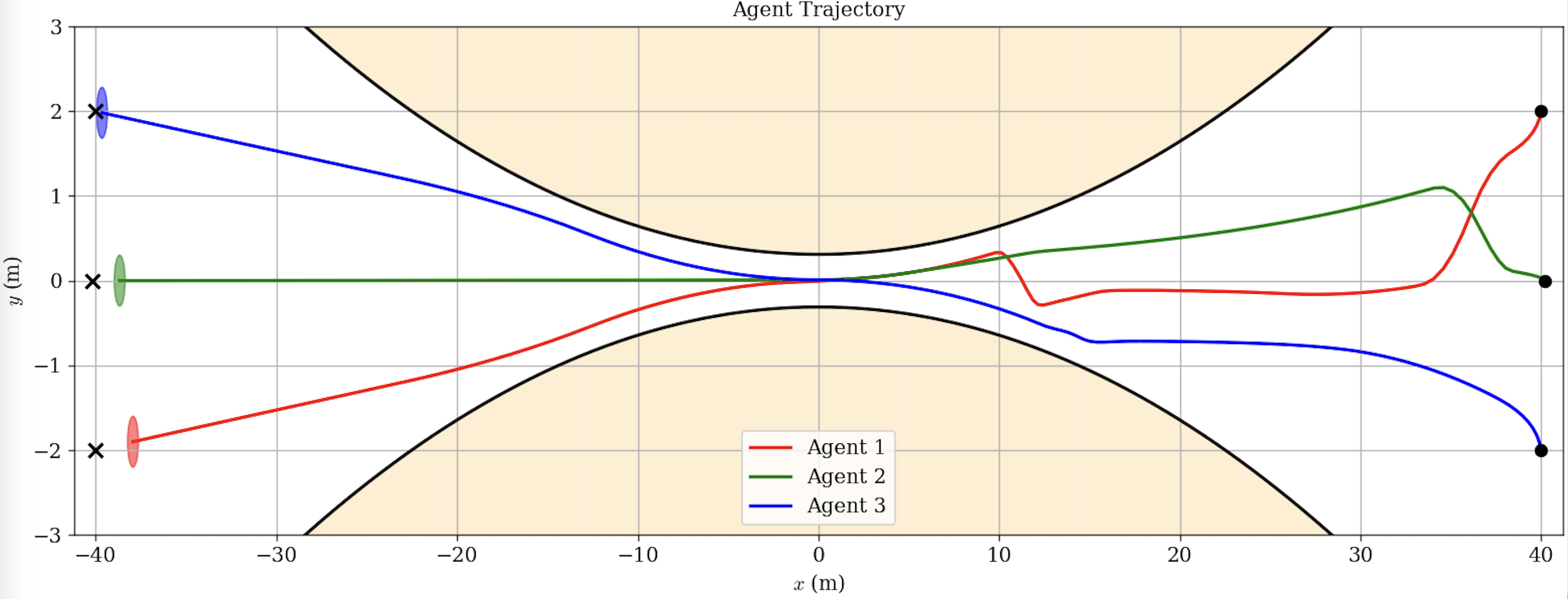}
        \caption{Goal reaching}
        \label{fig:goal_reaching}
    \end{subfigure}

    \caption{Motion snapshots of the landing-coordination scenario. Here agent 3 has the earliest landing time, followed by agents 2 and 1.}
    \label{fig:four_snapshots}
\end{figure}

Fig. \ref{fig:four_snapshots} highlights this landing coordination problem, where multiple aerial agents (3 agents here) share a common corridor (tunnel region between two orange semicircular arcs) and must satisfy prescribed landing-time constraints to reach their targets, marked by `x'.
The agents' motion cannot be planned independently as access to the landing strip must be coordinated. At the same time, different operators may value trajectory attributes differently, such as fuel use, control effort, or deviation from a preferred approach path. We model these private preferences through quadratic state and control penalties $Q_i$ and $R_i$. Before the landing sequence begins, each agent reports its preference parameters to a central planner, which allocates a collision-free corridor, and assigns a payment/transfer for the reported preferences.

The central planner returns to each agent its allocated corridor and transfer. Note, the transfer in this context can be interpreted either as a monetary credit/debit or as an adjustment to future landing preference. 
Each agent then plans its motion within the assigned airspace corridor. Hence, the corridor allocation is performed centrally so that each decentralized agent can independently plan its own path within the airspace without needing to account for collision constraints or any other agents' trajectory.

\subsection{Agent model and simulation parameters}

Each aircraft is modeled as a point-mass with a safety radius ($r=0.3$). For the current AAM planning problem, we consider horizontal-plane motion with double-integrator dynamics.
In the implementation, the state is $x_i=[p_x,p_y,v_x,v_y]^\top$, the control is planar acceleration, and the step size $dt =0.1 s$. The velocity ($v_x, v_y$) bounds are ([-10.5,10.5] $m/s$) in each component, and the control bounds are $[-22.5,22.5]$ units in each component.
\subsection{Nominal simulation and scalability}

We first show a nominal simulation to illustrate the basic landing-coordination behavior. For the three-agent scenario,
$x_{1,0}=[40,2,0,0]^\top$,
$x_{2,0}=[40,0,0,0]^\top$, and
$x_{3,0}=[40,-2,0,0]^\top$, with goals
$x_{1,\mathrm{goal}}=[-40,-2,0,0]^\top$,
$x_{2,\mathrm{goal}}=[-40,0,0,0]^\top$, and
$x_{3,\mathrm{goal}}=[-40,2,0,0]^\top$.
The nominal cost weights are
$Q_i=\mathrm{diag}(22.5,22.5,60,60)$ and
$R_i=\mathrm{diag}(1,1)$ for all $i\in\{1,2,3\}$.
The landing times specified for the agents are, $T_{\textrm{landing,1}}=22s$, $T_{\textrm{landing,2}}=20s$, and $T_{\textrm{landing,3}}=18s$ respectively.
We consider the case where agent 3 misreports its preference in this simulation,  $\hat{Q}_3= \lambda Q_3$ and $\hat{R}_3= \lambda R_3$, and misreporting factor $\lambda= 10^2$.
\begin{figure}[t] \centering \includegraphics[width=0.8\columnwidth]{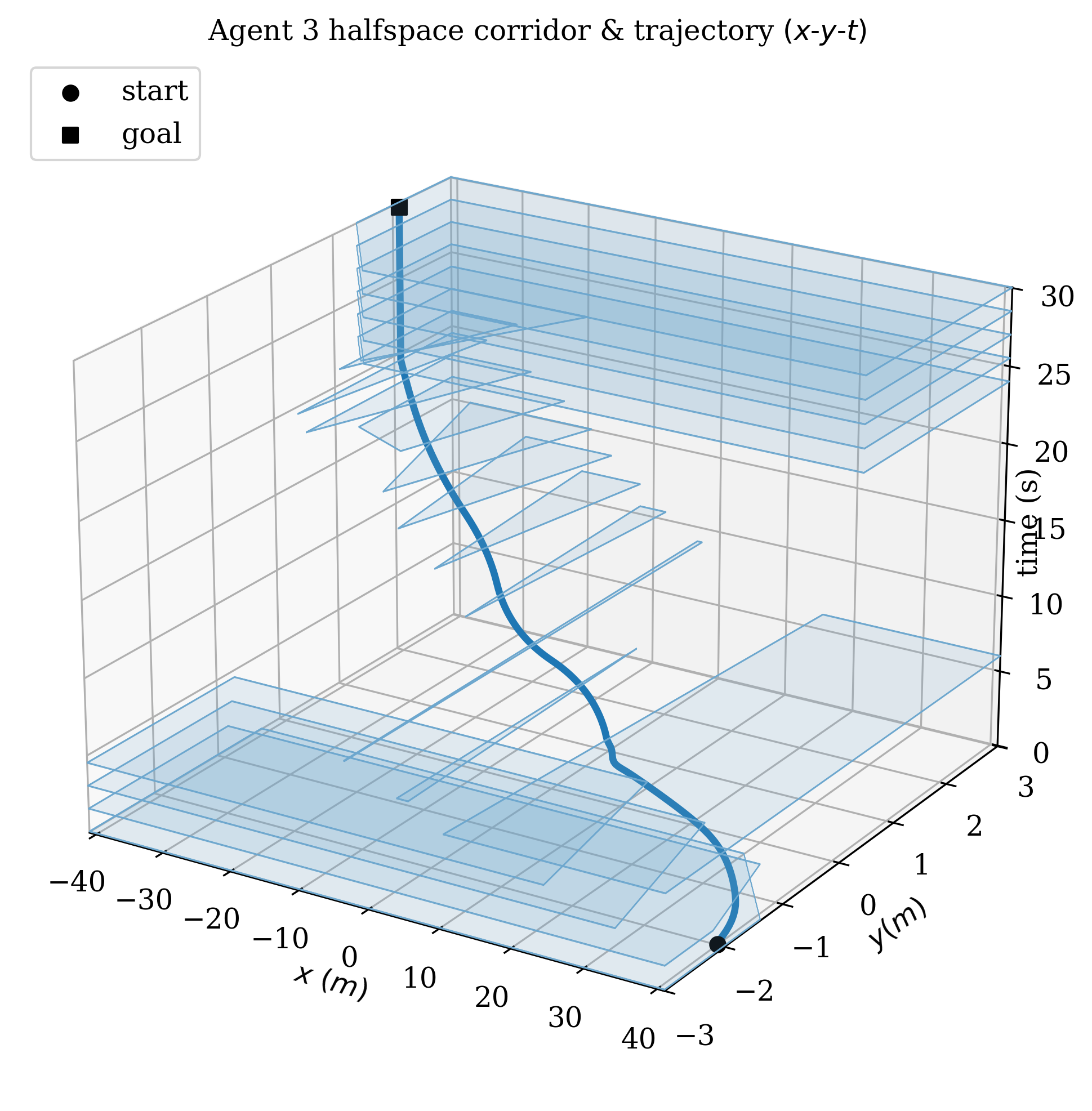} \caption{Allocated time-varying polytopic corridor of agent $3$. } \label{fig:halfspace} 
\end{figure}
Fig. \ref{fig:halfspace} shows the allocated airspace corridor along with the agent 3's trajectory as a function of time.
\begin{figure}[t] \centering \includegraphics[width=0.8\columnwidth]{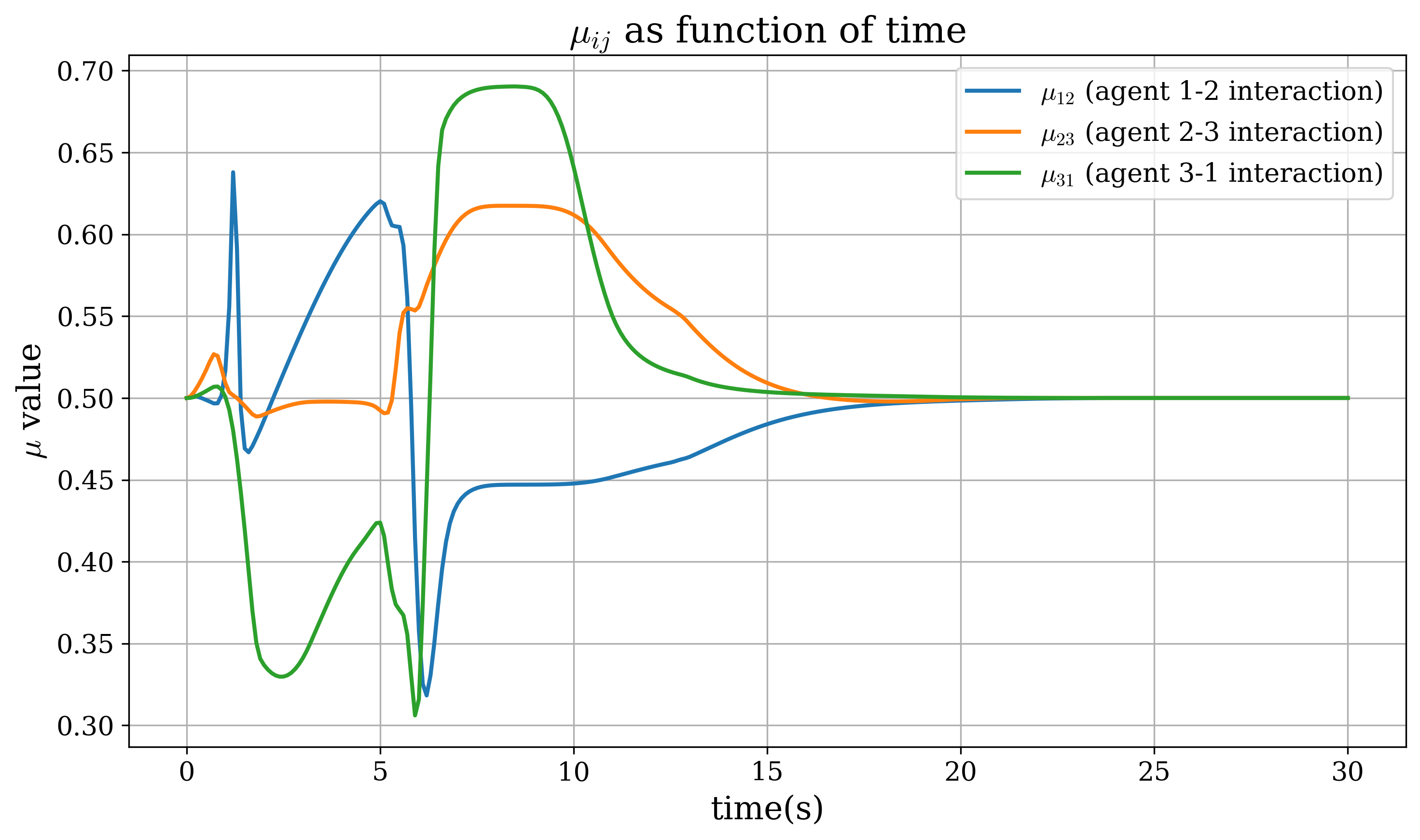} \caption{Halfspace mixing variables $\mu_{ij}$ as a function of time for different pairs of agents.} \label{fig:muplot} 
\end{figure}
Fig. \ref{fig:muplot} shows the halfspace mixing variables as a function of time.

Next, we perform a scalability study to analyze computation time at the
central and local levels. The all-agent allocation requires one nominal solve,
$N$ preliminary corridor-allocation solves, and one final reconciled solve.
Repeating the corresponding allocation with each agent removed for the
Clarke-pivot terms gives a total of $ N^2+2N+2 $
central nonlinear optimization calls. The number of central calls therefore
grows quadratically with $N$; however, this does not imply quadratic
wall-clock complexity, since the size and number of constraints of the
individual optimization problems also increase with $N$.
Table~\ref{tab:scaling_results} reports the central and agent-level computation
times for $N=3,4,5$. The results indicate that centralized computation is the
primary bottleneck, while the agent-level computation remains nearly constant
over the tested range.

\begin{table}[t]
\centering
\caption{Scalability study for the nominal landing problem}
\label{tab:scaling_results}
\begin{tabular}{c c c}
\hline
$N$ &  central time (s) & agent time (s) \\
\hline
3  & 147.08 & 2.26 \\
4 & 347.4 & 2.30 \\
5 & 924 & 2.34 \\
\hline
\end{tabular}
\end{table}

\subsection{Ablation study}
To evaluate the impact of various stages of our multistage central allocation pipeline on the system performance, from the perspectives of social welfare and truthfulness, we consider four variants of the planning architecture.

\textbf{P1: Full pipeline without VCG transfer.}
This variant uses the full multistage planning pipeline, including corridor allocation and decentralized execution, but no VCG transaction is applied. It serves as a baseline to isolate the role of the incentive mechanism.

\textbf{P2: Stage-1-only exact centralized planning.}
In this case, only Stage 1 is used. The planner solves the exact centralized collision-avoidance problem, computes the VCG payment from that solution, and sends the resulting payment and reference trajectory to the agents. However, no corridor information is shared. Thus, the agents solve for control inputs to simply follow the assigned reference trajectory and do not receive an allocated halfspace/corridor description.

\textbf{P3: Fixed corridor allocation.}
This variant uses Stage 1 followed by a fixed second-stage halfspace allocation with
\begin{equation}
\mu_{ij,k}=0.5,\qquad \forall i\neq j,;k.
\end{equation}
The corridors are constructed for the agents using this fixed split and then shared with them. Each agent then solves its local optimal control problem subject to remaining within its assigned corridor.

\textbf{P4: Proposed multistage corridor allocation.}
This is the full proposed method. Stage 1 generates the nominal trajectories, Stage 2 optimizes the corridor allocation through $\mu$, VCG-inspired transfers are computed from the resulting allocation, and the agents then solve decentralized local planning problem (\ref{agent1}) within the assigned corridors.

\subsubsection{Monte Carlo experiment setup}

To assess the performance of $P4$ and compare against the baselines, we perform Monte Carlo experiments over admissible initial conditions and agent preference profiles. We conduct $70$ simulation runs, for which we randomly sample the initial \(x\)-position from \([30,50]\) m, the goal \(x\)-position from \([-50,-30]\) m, the initial and goal \(y\)-positions from \([-4,4]\) m, the initial velocity components \(v_x\) and \(v_y\) from \([-0.5,0.5]\) m/s, the cost matrix scaling factor from \([1,50]\), and the misreporting agent index from \(\{1, 2, 3\}\). Note, the sampled initial and final positions are chosen to be within the free space. Across these trials, we evaluate both efficiency and incentive behavior. Moreover, for each simulation run, we carry out two studies: $1)$ all agents are truthful, $2)$ one randomly selected agent misreports its cost matrices by a factor of $100$ to gain an unfair advantage.

\subsubsection{Social valuation analysis}
First, we compare the methods P2, P3, and P4 based on social-welfare performance. (P1 does not involve any transfer, so we do not consider P1 for this analysis). The social valuation is defined as follows \cite{chen2025two},
\begin{equation}
V_{\mathrm{soc}}
=
-\sum_{i=1}^{N} J_i(X_i,U_i,\theta_i),
\end{equation}
\begin{table}[t]
\centering
\caption{Average Social Valuations ($\times 10^8$) Across Monte Carlo Trials (higher valuation is desired)}
\label{tab:social_val}
\begin{tabular}{c c c c}
\hline
Reporting &$P2$ &  $P3$ & \textbf{P4 (Ours)} \\
\hline
Truthful  & -3.30854 & -3.30907&\textbf{-3.30782} \\
Misreport &-3.3431 &-3.34105 &\textbf{-3.33867} \\
\hline
\end{tabular}
\end{table}

From the Monte Carlo studies carried out for both the truthful and misreporting cases, as shown in Table \ref{tab:social_val}, we can infer that the social valuation is comparatively higher in our method compared to the baselines P2 and P3.

\subsubsection{Study on the role of VCG transfer}

We now compare truthful and strategic misreporting for P1 and P4 (recall, P1 setup is the same as P4, except P1 does not have any VCG transfer associated with agents). We examine the agent's \textit{utility} (\ref{utileq}).
 We define the difference between the utility of truthful and misreported cases as,
\begin{equation}
\Delta u_i=
u_i^{\mathrm{truth}}-u_i^{\mathrm{misreport}}.
\end{equation}
For a strategy-proof mechanism, one expects  $\Delta u_i \ge 0$
since truthful reporting should weakly dominate misreporting and should be incentivized. 

In our Monte Carlo runs, we found the mean $\Delta u$  of P1 to be $-2.4\times10^5$, whereas the mean $\Delta u$ for our method P4 is $2.74\times 10^6$. This implies that adding VCG-based payments helps incentivize truthful reporting.
In our analysis, we observe that in some Monte Carlo runs for P4, the $\Delta u$ takes negative values. This is because our approach (P4) is a multistage decomposition, and hence only an approximation, of the original (strategyproof) central allocation problem (\ref{eq:central1}). Hence, our proposed solution is only incentive-aware, since $\Delta u$ is greater than $0$ in our case, on average sense.

\section{Conclusions and Future Work}

This paper proposed an incentive-aware framework for multi-agent collision
avoidance in shared AAM airspace with privately known agent preferences. By
combining convex corridor allocation, a VCG-inspired transfer rule, and a
multistage decomposition of a bilevel optimization, the framework provides a
tractable approach to centralized airspace allocation with decentralized
agent-level execution.

The numerical results demonstrate safe decentralized execution and improved
social valuation relative to the considered ablation baselines. The
VCG-inspired transfer also increases the average truth-minus-misreport utility
relative to the no-transfer case. However, negative values of $\Delta u$ occur
in some Monte Carlo trials because the multistage procedure provides only a
candidate solution to~\eqref{eq:central1}, rather than the exact VCG
allocation. The implemented mechanism is therefore best interpreted as
incentive-aware rather than strategyproof.

The present study uses deterministic planar double-integrator dynamics, does
not provide a formal approximation or incentive-loss bound, and retains
centralized leave-one-out computations as the main scalability bottleneck.
Future work will investigate iterative or more scalable allocation schemes,
stronger theoretical guarantees for the approximate mechanism,
higher-fidelity dynamics, uncertainty, communication delays, and dynamic
obstacles through uncertainty-aware or receding-horizon corridor allocation,
as well as repeated-allocation settings in which transfers may represent
future landing priority or access credits.

\bibliography{main}
\end{document}